\documentstyle[11pt,epsfig]{article}
\author{M. Neek-Amal$^1$,  R. Moussavi$^1$ and H. R.
Sepangi$^{1,2}$\\  {\small $^1$Computational Physical
Sciences Research Laboratory, Department of Nano-Science,}\\
{\small Institute for Studies in Theoretical Physics and
Mathematics (IPM), P.O. Box 19395- 5531,Tehran, Iran.}\\{\small
$^2$Department of Physics, Shahid Beheshti University, Evin,
Tehran 19839, Iran}}
\title{\bf Monte Carlo simulation of size-effects on thermal
conductivity in a 2-dimensional Ising system}
\begin{document}
\maketitle \maketitle \vspace{0.5cm}
\begin{abstract}
A model based on microcanonical  Monte Carlo method is used to
study the application of the temperature gradient along a
two-dimensional (2D) Ising system. We estimate the system size
effects on thermal conductivity, $K$, for a nano-scale Ising layer
with variable size. It is shown that $K$ scales with size as $\
K=cL^\alpha$ where $\alpha$ varies with temperature. Both the
Metropolis and Cruetz algorithms have been used to establish the
temperature gradient. Further results show that the average demon
energy in the presence of an external magnetic field is zero for
low temperatures,  \vspace{0.5cm}
\end{abstract}
\pagebreak
\section{{  Introduction}}
The Ising system was introduced many decades  ago to study simple
spin systems. Over the years it has stood the test of time and
proved to be useful in many areas of physics, from statistical
mechanics to biological systems $\cite{nelson}$. As such, it has
also been in use to study large scale Monte-Carlo (MC) simulations
of complex systems ever since the advent of adequate computing
power to tackle such challenges. As is well known, the usual MC
algorithm is a useful approximation if the time scales over which
thermal diffusion occurs are very short. It is therefore clearly
inadequate for problems where the dynamical process is controlled
by local variations in temperature, including the calculation of
thermal conductivity. To avoid such shortcomings, Creutz
\cite{crtz} developed a micro-canonical algorithm by introducing
``demons'' to control distribution of energy throughout the
system. Such demons may be conceived to act as thermometers. This
is particularly helpful in  performing non-equilibrium simulations
with unequal local temperatures. This development was the basis of
thermal conductivity computations in a 2D Ising system presented
in \cite{harris}. The same method was further developed later by
Mak \cite{mak} to accommodate external magnetic fields.

In this paper we use the Creutz algorithm together with Fourier
law to calculate thermal conductivity within the framework of a 2D
Ising system. In doing so, we have also investigated the size
effects and shown that thermal conductivity scales as $\ K=c
L^\alpha$ where $\alpha$ varies with temperature and $L$ is the
size of the system. We further show that the average demon energy
when an external magnetic field is present is zero at low
temperatures. Also, the demon thermometer, as developed by Mak,
leads to unacceptable temperature gradients. However, this
thermometer works satisfactorily at high temperatures.
\section{Ising model and thermal conductivity}
Let us first briefly consider the salient features of the Ising
system in 2D relevant to our present discussion. This model is
characterized by the interaction Hamiltonian $H$ given by
\begin{equation}
H=\frac{J}{2} \sum_i\sum_{j\ne i}{S_i}{S_j}\,- \,h\sum_i{S_i},
\label{eqn3}
\end{equation}
where the spin variables $\ S_i$ take values from the set
$\{1,-1\}$ with $\frac{1}{2}$  denoting the set of all nearest
neighbor pairs and $h$ is proportional to a uniform external
magnetic field. The exchange constant $J$ is a measure of the
strength of the interaction between nearest neighbor spins. If
$J<0$, we expect the state of the lowest total energy to be
ferromagnetic, {\it i.e.} the spins all point in the same
direction. For $J>0$, the state of the lowest energy is expected
to be anti ferromagnetic, that is, alternate spins are aligned.
Here we concentrate on the ferromagnetic case. A particular
configuration of microstates of the lattice is specified by the
set of variables $\{S_1,S_2,\cdots,S_n\}$ for all lattice sites.
We work with units in which $1/T=\beta$.
\subsection{{\bf Thermal conductivity from
Monte Carlo simulations }} The method we have used is based on the
micro-canonical algorithm of Creutz which is complementary to the
standard Monte Carlo method. Here, we investigated the transport
properties in a non-equilibrium steady state. Let us first note
that as temperature is an incoming parameter to the metropolis
algorithm, having been fixed before hand, the system changes with
a fixed temperature and there would be no possibility to explore
the behavior of the system if the temperature changes from one
part of the system to the other. However, if we were to study a
system with local temperature variations, we would be able to
compute such transport coefficients as thermal conductivity.
Creutz has shown that the method can also be used for dynamical
properties, in a qualitative study of thermal conductivity
\cite{crtz}.

To this and other ends, we first divide our 2D system and make it
into many sub-layers. We then equilibrate the system (all layers)
at temperature $T_m$ using Metropolis algorithm and as usual,
periodic boundary conditions are used along each direction. After
equilibration we select a few layers from each end and fix their
temperatures at $T_{h}$ ($h$ for high) and $T_l$ ($l$ for low)
using Metropolis algorithm. At this point we have to remove
periodic boundary condition in the $x$-direction, {\it i.e.} along
the direction for which a temperature gradient is expected. We
apply periodic boundary conditions along the direction for which
there is no temperature gradient, that is, the $y$-direction. Also
note that $T_m=\frac{T_h+T_l}{2}$ and the symbol $T$ instead of
$T_m$ has been used in all the figures. After many iterations we
can calculate thermal conductivity of the system at $T_m$ using
the Fourier's law for heat conduction $\cite{Allen}$. The
temperature gradient is then deduced from the resulting linear
temperature profile, that is, from the Fourier law
\begin{equation}
{\bf Q}=-K{\bf \nabla} T ({\bf x},t),\label{eqn1}
\end{equation}
where ${\bf Q}$ is the heat current and $K$ is the thermal
conductivity of the system.

To study the behavior of a temperature gradient, we use the
``demon algorithm'' \cite{crtz}. To understand the mechanism
behind this algorithm  suppose  we add an extra degree of freedom
$E_d(l)$ to each original layer separately. These extra degrees of
freedom are called ``demons.'' The demons travel along each layer
transferring energy as they attempt to change the dynamical
variables of the layer. If a change lowers the energy of the layer
the excess energy is allocated to the demon. If the change raises
the energy of the layer, the demon transfers the required energy
to the layer if it has sufficient energy available. The only
constraint is that the demons cannot have negative energy. The
mean demon energy $\langle E_d(l)\rangle$ in the absence of an
external magnetic field is given by
\begin{equation}
{T}=\frac{4}{\log\left( 1+\frac{4}{\langle E_d(l)\rangle}\right)}.
\label{eqn6}
\end{equation}
The above equation was derived based on the fact that the
probability distribution of the demons for each layer is
Maxwellian. In many respects, the demon acts as a thermometer
since it has only one degree of freedom in comparison to the many
degrees of freedom of the system with which it exchanges energy.
After taking sufficient time steps during which the temperature of
the end layers were being monitored by the Metropolis method and
that of the mid-layers by the demon method, a stationary
temperature gradient begins to set in. To determine it
qualitatively, we calculate the ``local'' temperature with the use
of equation (\ref{eqn6}) for each sub-layer. In figure 1 we have
plotted the temperature of sub-layers versus the $x$-coordinates
of the layers corresponding to the original lattice for several
high and low temperatures. As can be seen, the fluctuations of the
temperature profiles about the mean temperature are extremely
small. Figure \ref{fig2} shows variation of the magnetization
profile versus the lattice size per spin. These figures clearly
show that a steady state in heat transport has set in. It is also
worth mentioning that as $T_m$ gets closer to the critical
temperature, a temperature gradient still persists. To find the
critical temperature $T_c$ in our system, we did a simulation
using a greater size (10000 spins in a $100\times 100$ lattice)
and found that the critical temperature is near 2.3 which is very
close to the exact result of $2.269$ \cite{crtem}, as shown in
figure \ref{fig3}. The exact value for $T_c$ has been used
throughout the paper for scaling the horizontal axes for all the
relevant figures. The heat flux $\bf Q$ is the amount of energy
that the hot layer adds to the system and is computed by
subtracting the total energy of the hot $(E_{hot})$ and cold
$(E_{cold})$ layers per unit area per Monte Carlo step for each
$T_m$ resulting from individual $T_{h}$ and $T_{l}$. For
calculating thermal conductivity we use
\begin{equation}
K(T_m)=\frac{E_{hot}-E_{cold}}{\frac{\Delta{T} } {\Delta{X}
}{A\tau} },\label{eqn7}
\end{equation}
where $A$ is the area which in our 2D model represents the system
size in the $y$-direction and $\tau$ is conventionally measured in
terms of the Monte Carlo steps per spin. It should be noted that
the effects of the mid-layers in the calculation of thermal
conductivity are contained in the  temperature gradient $\Delta
T/\Delta X$ which changes from temperature to temperature.
\section{Theoretical considerations}
In this section we shall be concerned with an analysis in which we
investigate the dependence of heat conductivity on the size of the
system and temperature. Our analysis is based on the equilibrium
kinetic Ising system and the Green-Kubo relation \cite{Green,Kubo}
for heat conductivity according to which
\begin{eqnarray}
K(T,L)=\frac{1}{3K_{\tiny\mbox{B}}T^2}\int_{0}^{\infty} dt\int
dr\langle{\bf Q}(0,0)\cdot {\bf Q}(r,t)\rangle,\label{eqn2}
\end{eqnarray}
where ${\bf Q}$ is now the instantaneous heat current vector in an
assembly of $N$ interacting spins, $T$ is the temperature and the
angle brackets, $\langle\cdots\rangle$, indicate equilibrium
ensemble average. Definition of $\bf Q$ in terms of the variation
of magnetization is an interesting question to which a
considerable amount of attention is being paid \cite{stat}. It
would be useful to note that the integrand in equation
(\ref{eqn2}), as will be discussed below, depends on the size of
the system and therefore integration on $r$ has to be done
independently.
\begin{figure}
\centerline{ \epsfig{figure=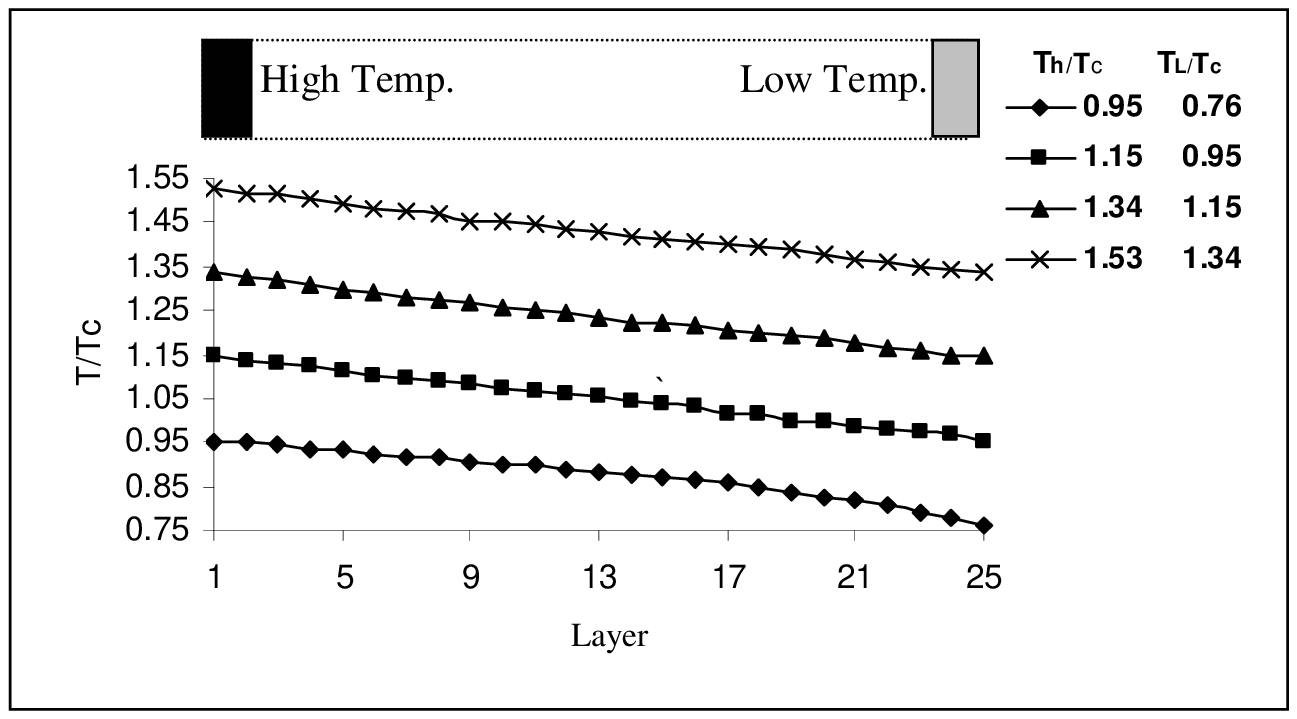,width=10cm}}
\caption{\footnotesize  Local temperature profile in a $100\times
100$ Ising lattice with different gradients.} \label{fig1}
\end{figure}

Heat current vectors depend on the rate of change of the direction
of spins. Consequently, transferring  this change to the direction
of temperature gradient is highly dependent on the correlation
functions. Also, correlation functions depend on temperature and
vanish above $T_c$,  their values approaching unity around $T_c$.
In fact, heat transport is the transmission of information by the
spins as they flip. In the limit of high temperature, heat
transport hardly occurs because both the correlation functions and
transmission of the flipping of the spins are small. However, each
spin flips independently of the others and takes an amount of
energy equal to $\pm 2J$. This means that the size of the system
becomes an irrelevant parameter, so that thermal conductivity
becomes independent of the size. In view of the above discussion,
we estimate the integral as
\begin{eqnarray}
K(T)\sim\frac{J^2}{T^2}, \label{eq10}
\end{eqnarray}
where $J$ is independent of temperature. In addition, in the limit
of low temperature and near $T_c$, correlation functions increase
in value and  transmission of information by the spins being
flipped is large whereas in the limit $T\rightarrow 0$ the
transmsion would be zero, leading to a zero thermal conductivity.
When correlation length is larger than the system size, the
boundary of the system influences heat transport since the
characteristic wavelength of the system is smaller than the
correlation length. We expect thermal conductivity to increase as
the system size grows. This increasing regime finally stops when
system size becomes greater than the correlation length. In the
low temperature limit however, each spin has a small chance of
being flipped and near zero takes an amount of energy equal to
$\epsilon=8J$ with probability $\exp(-\epsilon/T)$ after a
flipping has occurred. Indeed, near $T_c$, $\epsilon$ is smaller
than $8J$. Summarizing, we may deduce the following formula for
thermal conductivity
\begin{eqnarray}
K(T,L)\sim\frac{J^2L^\alpha\exp(-\epsilon/T)}{T^2}, \label{eq11}
\end{eqnarray}
where $\alpha$ is a dimensionless positive quantity that tends to
zero in the limit of large system sizes. The dependence of $K$ on
$L$ may be better understood in the light of the dependence of
specific heat $C_B$ on the system size and its relation to thermal
conductivity which may be crudely written as $K\sim C_BL$. It
would be interesting to note that the same results have been
reported in \cite{lep} for various potential models.
\section{{\ Results}}
A discussion of the results we have obtained from our simulations
is now in order. The initial configuration of the 2D system was
selected by randomizing square lattices consisting of $N$=1024,
2025, 3247, 4761, 6400, 8100 and 10000 spins. The lattice was
layered along the $x$-axis, each layer consisting of 64, 125,
$\cdots$ sites respectively and may also consist of 2, 3, or 4
columns of spins in the y direction. A typical configuration
includes $100\times 100$ cites consisting of 25 layers, each made
of 4 columns of spins. For simplicity, we adopt $J$ as the unit of
energy.
\begin{figure}
\centerline{ \epsfig{figure=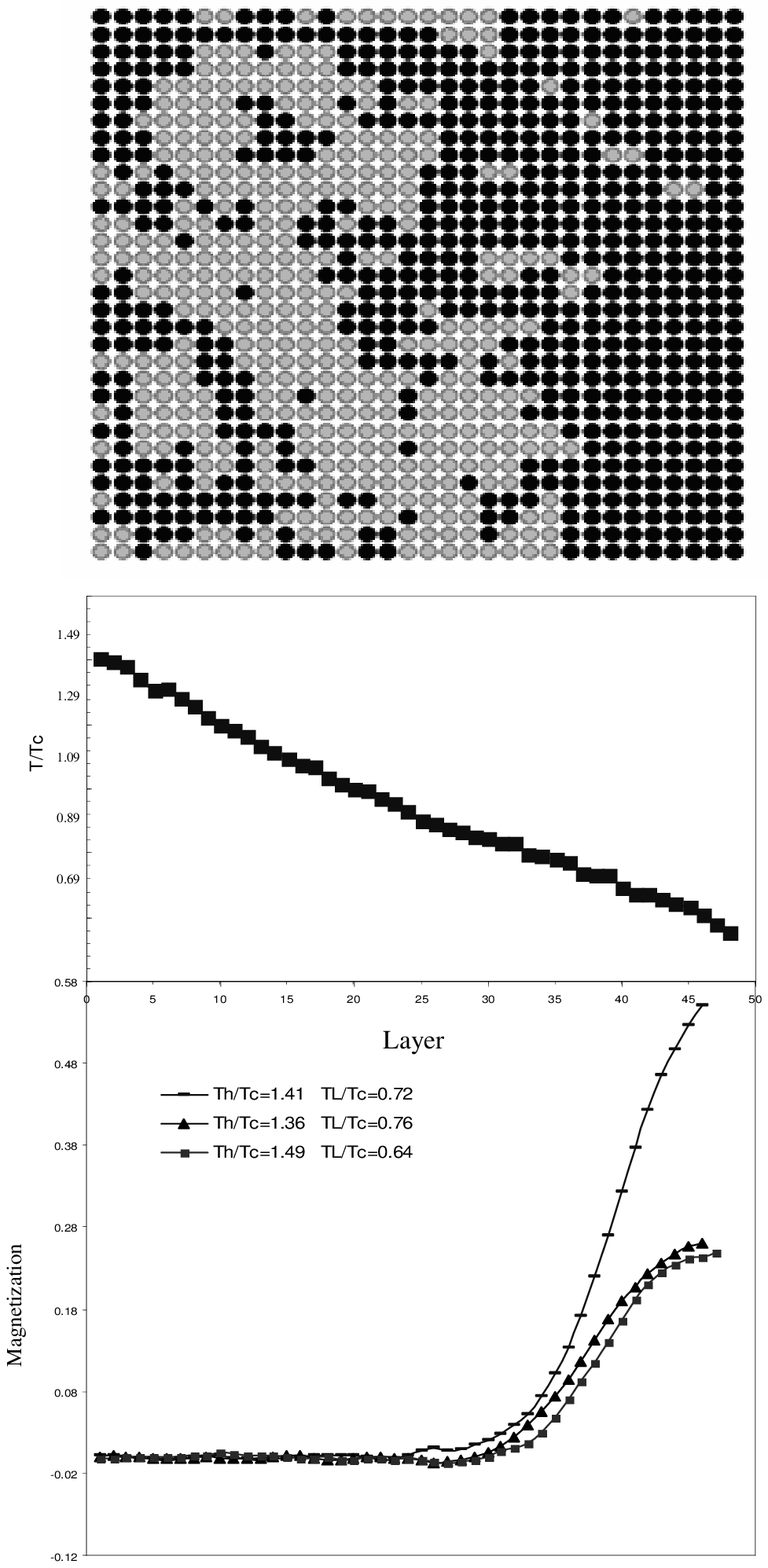,width=10cm}}
\caption{\footnotesize Variation of the magnetization profile per
spin versus the lattice size for three different temperature
gradients with a lattice size of $100\times 100$. The figure in
the middle corresponds to the largest temperature gradient and
that on the top is a snapshot of a system consisting of $32\times
32$ spins with the same temperature gradient as the figure in the
middle. The dark and light points represent opposite spins. Units
are arbitrary.} \label{fig2}
\end{figure}

Thermal conductivity in each case for various sizes has a peak
around the critical temperature and shows the usual qualitative
behavior for many materials $\cite{thermal}$ (metals and alloys)
as shown in figure 4. It can be seen that as temperature
decreases, conductivity of the system increases. The reason is
that in the limit of low temperature, correlation length for the
spins increases and therefore, as soon as a small amount of heat
enters from one side, it would be conducted easily through the
spins. This means that the coefficient of conductivity becomes
large. Also, when temperature approaches that of the critical
$(\approx 0.8T_c)$, $K$ has a maximum. In the limit of high
temperature, thermal conductivity is consistent with the
$\frac{1}{T^2}$ behavior, as we expect from the above theoretical
discussion. As a matter of discussion, it should be mentioned that
one cannot express thermal conductivity in the Ising system in any
quantitative way because we employ a rather simple model and do
not know anything about the time and distance between adjacent
spins in our Monte Carlo simulations. Therefore, one cannot say
much about the unit of thermal conductivity. Equation (\ref{eq11})
shows that in the limit of low temperature, $K$ passes through a
maximum in the Ising system. We have fitted our simulation data
using this equation and obtained $\epsilon=1.45$ for ferromagnetic
and $\epsilon=1.47$ for anti-ferromagnetic  cases as has been
shown in figure 5. It is worth mentioning at this point that the
values for $\epsilon$ mentioned above have been obtained using the
interval in which thermal conductivity is peaked. Furthermore, in
the high temperature limit the correlation length and correlation
time are small and the Green-Kubo relation leads to equation
(\ref{eq10}) which is in good agreement with our simulation
results, see figure 6. As can be seen in figures 4 and 6, our
calculations are in agreement with both ranges of temperatures. We
have also demonstrated, within the context of this model, that for
a given temperature gradient, thermal conductivity increases with
size as
\begin{equation}
K=cL^\alpha, \label{eqn8}
\end{equation}
where $\alpha$ is a positive function of temperature and
approaches zero at high temperature, as is shown in the table
attached to figure 7 which exhibits thermal conductivity
decreasing as the lattice size is reduced. This is an interesting
behavior  and is in agreement with the above discussion on size
effects.
\begin{figure}
\centerline{ \epsfig{figure=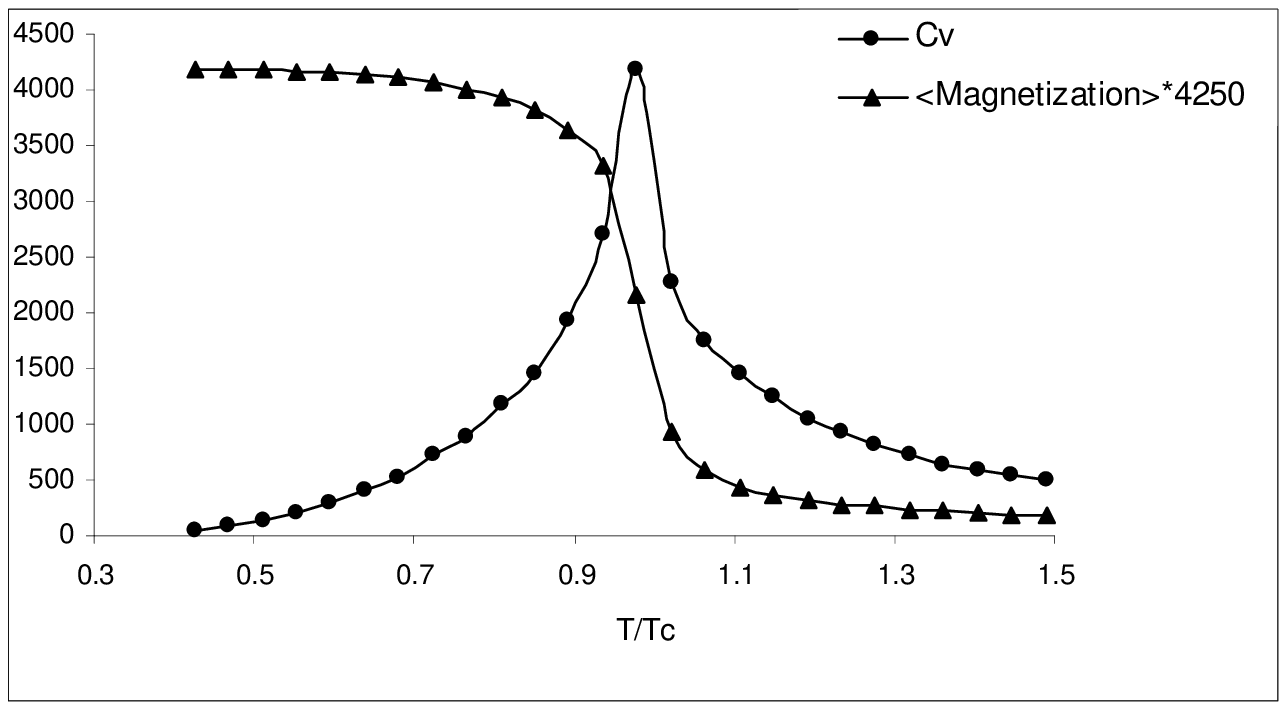,width=10cm}}
\caption{\footnotesize  This figure shows the location of the
critical temperature, $T_c$. In our simulation, it happens at
T=2.3 for a $100\times 100$ lattice. Horizontal axis is scaled by
$T_c=2.269$, the exact value for $T_c$. Units are arbitrary.}
\label{fig3}
\end{figure}

In the presence of an external magnetic field $h$, equation
(\ref{eqn6}) would be converted to the following equation
\begin{equation}
{T}=\frac{4/d}{\log\left( 1+\frac{4/d}{\langle
E_d(l)\rangle}\right)}, \label{eqn66}
\end{equation}
where $d=\frac{2b}{h}$ with $b$ and $d$ being relative prime
positive integers and, in the limit of $h\rightarrow 0$, we
recover equation (\ref{eqn6}) as has been described in \cite{mak}.
In the limit of zero temperature or near the critical temperature,
due to the freezing of spins, the demon energy for each layer,
$\langle E_d\rangle$, would be reduced in such a manner as to
result in a zero average. This can be investigated by letting
$\langle E_d\rangle$ go to zero in equation (\ref{eqn66}) which
will result in unacceptable values for local temperatures. Even in
limit $T/T_c <2$ we could not obtain good results for temperature
gradients, in contrast to  higher temperatures where reasonable
gradients are obtained similar to those shown in figure 1.
\begin{figure}
\centerline{ \epsfig{figure=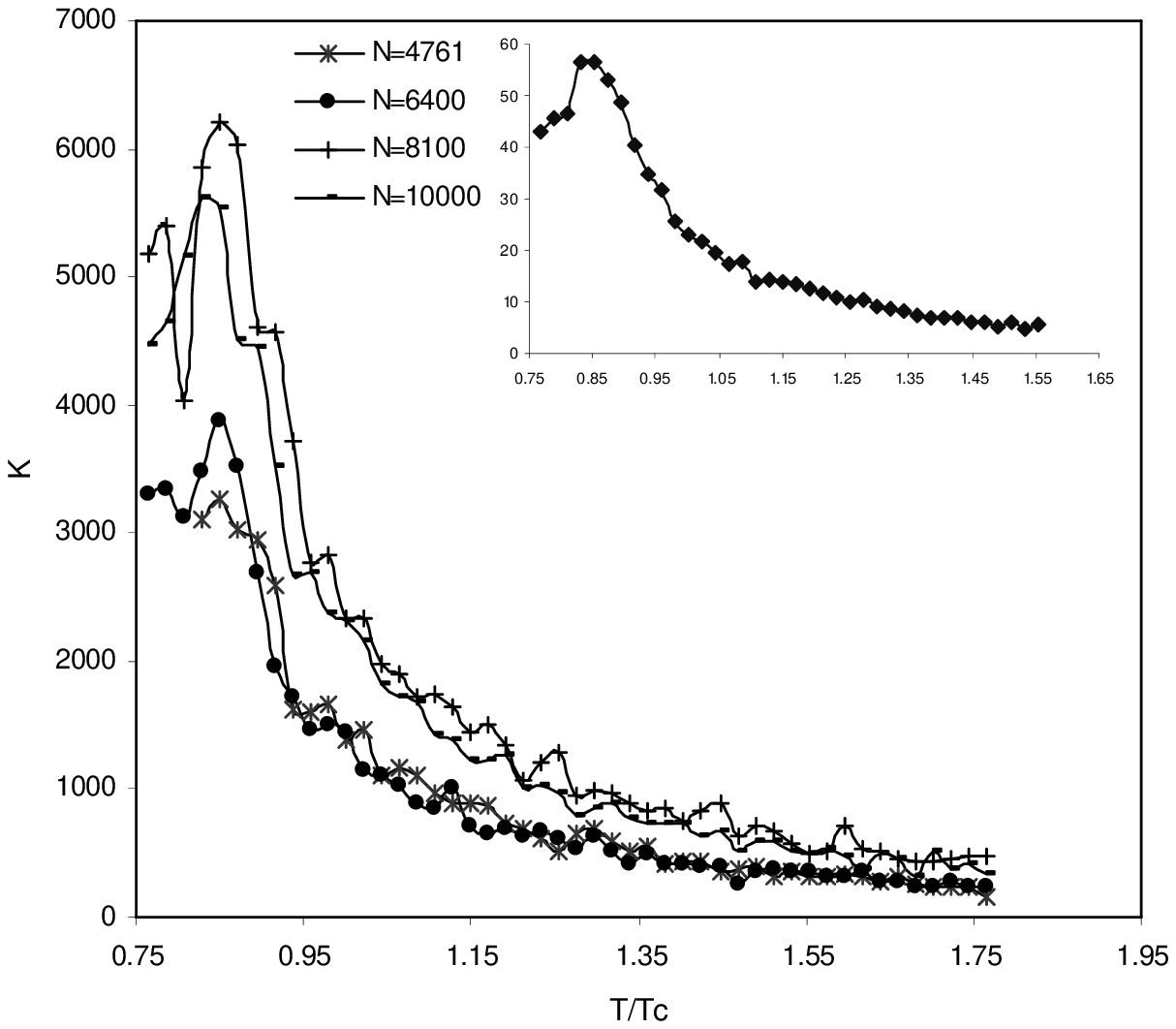,width=10cm}}
\caption{\footnotesize  Temperature dependence of thermal
conductivity for various system sizes. This figure shows that for
temperatures less than $T_c$, $K$ increases and passes through a
maximum. Afterwards, $K$ would decrease according to equations
(6,7). Note  that as the system size becomes larger, the peaks
representing $T/T_c$ approach unity. The insert indicates the same
results for a $100\times 100$ antiferromagnetism system. Units are
arbitrary.} \label{fig4}
\end{figure}
\begin{figure}
\centerline{ \epsfig{figure=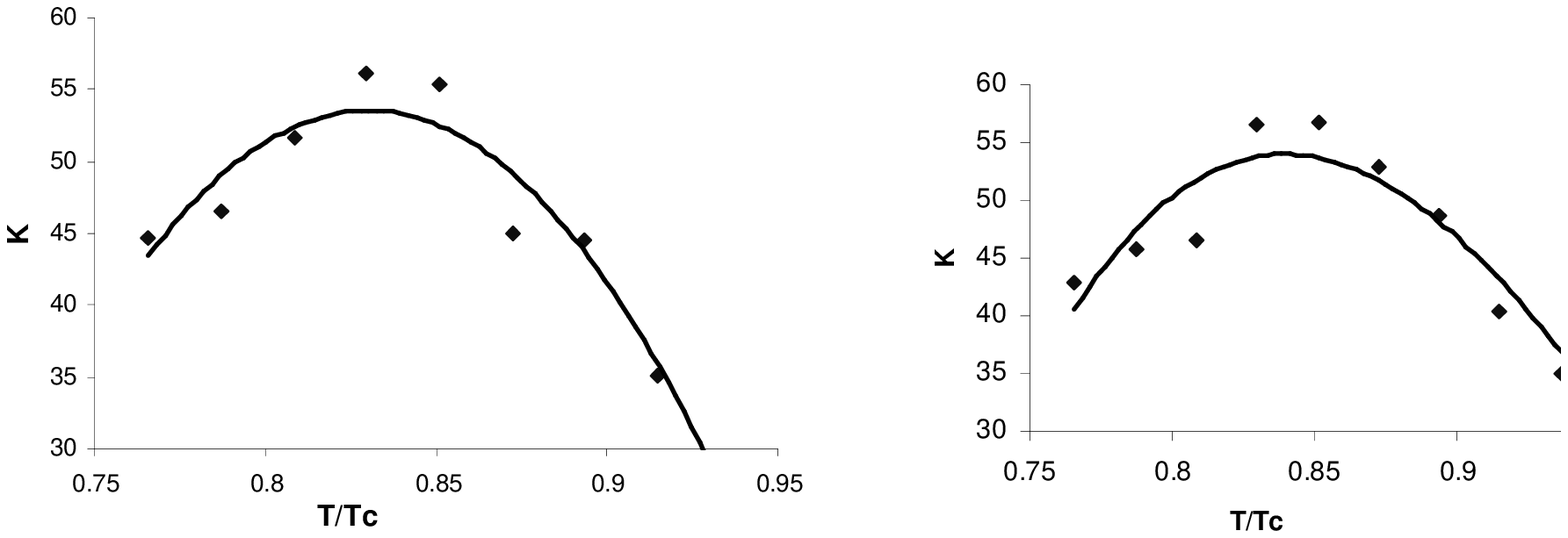,width=15cm}}
\caption{\footnotesize  Points showing simulation results for
temperatures smaller than $T_c$ and around it. The solid line
demonstrates the behavior according to equation (7), left, for
$J<0$   and right,  for $J>0$ with arbitrary units.} \label{fig5}
\end{figure}
\begin{figure}
\centerline{ \epsfig{figure=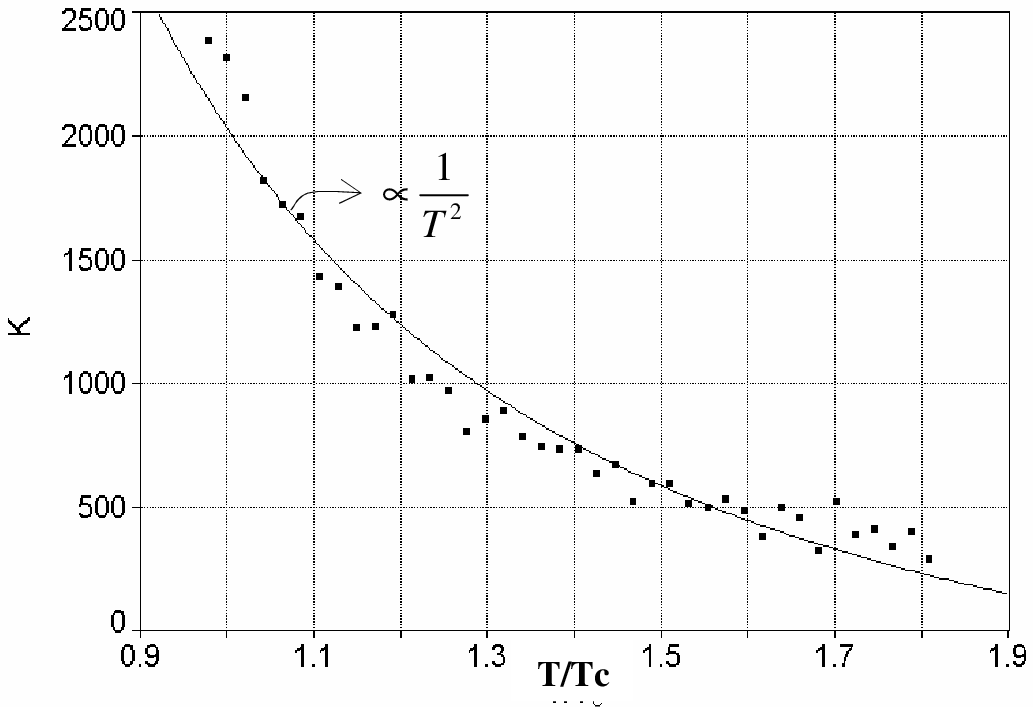,width=10cm}}
\caption{\footnotesize  The points present simulation results and
the solid line showing theoretical prediction for temperatures
above $T_c$ according to equation (6) with arbitrary units.}
\label{fig6}
\end{figure}
\begin{figure}
\centerline{ \epsfig{figure=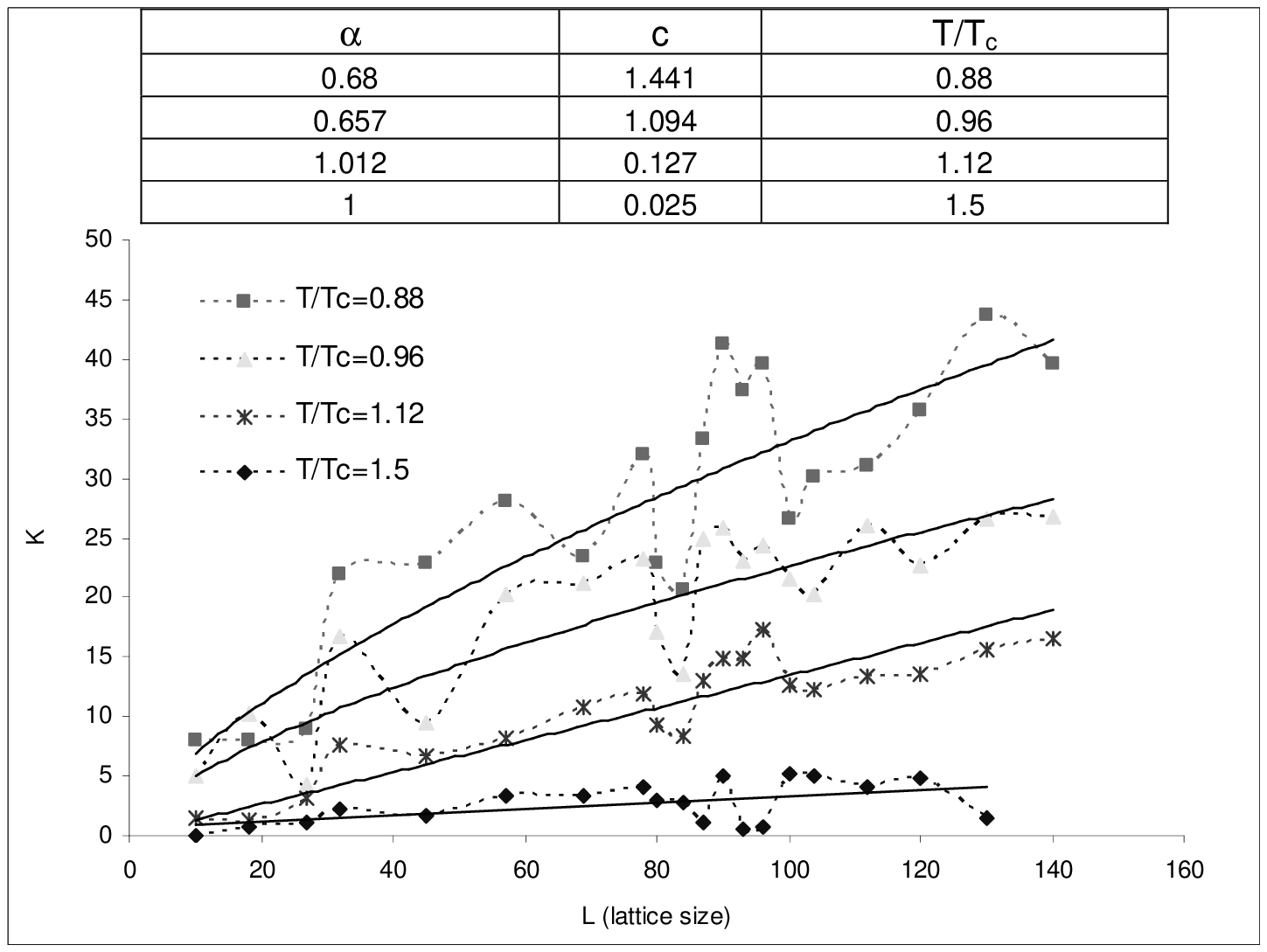,width=10cm}}
\caption{\footnotesize  The effect of size on thermal
conductivity. $K$ grows from zero exponentially and, as can be
seen, decreases with temperature. The table shows how the values
of $c$ and $\alpha$ change with temperature in equation (8) with
arbitrary units.} \label{fig7}
\end{figure}
\section{Conclusions}
In spite of its simplicity, the Ising system has been used in a
myriad of applications in modelling and simulating various
systems. Figures 4 and 7 make it easy to see the size effects
becoming important in the small spin systems within the context of
the model presented here. These figures illustrate the influence
of both the size and temperature effects on  thermal conductivity
for a simple 2D Ising system. The influence of temperature,
however, is not a new subject theoretically. These figures show
that thermal conductivity increases when temperature is reduced.
On the other hand, figure 7 shows that thermal conductivity would
increase exponentially with increasing system size and goes to
zero with a decreasing size in any temperature.

Although our study in this paper has been based on a 2D model, the
results can be generalized using a 3-dimensional
lattice.\vspace{10mm}\\
{\bf Acknowledgement}\vspace{2mm}\noindent\\
We would like to thank H. Rafii-Tabar for useful discussions.

\end{document}